\documentclass[12pt]{amsart}

\usepackage{amscd}
\usepackage{mathrsfs}
\newtheorem{theorem}{Theorem}[section]
\newtheorem{lemma}[theorem]{Lemma}

\theoremstyle{definition}

\newtheorem{remark}[theorem]{Remark}
\numberwithin{equation}{section}

\def\H{\mathscr H}
\def\K{\mathscr K}
\def\R{\mathscr R}
\def\N{\mathscr N}
\def\D{\mathscr D}
\def\G{\mathcal G}
\def\uno{\mathsf 1}
\def\zero{\mathsf 0}
\def\E{\mathsf {E}}
\def\RE{\mathbb R}
\def\CO{{\mathbb C}}

\def\fh{\mathfrak h}

\begin{document}

\title[On the common Point Spectrum of Self-Adjoint Extensions]
{On the common Point Spectrum of pairs of Self-Adjoint Extensions}

\author{Andrea Posilicano}
\address{DiSAT - Sezione di Matematica,  Universit\`a
dell'Insubria, I-22100 Como, Italy}

\email{posilicano@uninsubria.it}

\begin{abstract} Given two different self-adjoint extensions of the same symmetric operator, we analyse the intersection of their point spectra. Some simple examples are provided.
\end{abstract}

{\maketitle }
%\vskip 10pt
{\em  \small Dedicated to Vladimir  Koshmanenko on the occasion of his 70th birthday}
%\vskip 20pt
\begin{section}{Preliminaries}
Given a linear closed operator $L$, 
we denote by
$$\D(L)\,,\quad\K(L)\,,\quad\R(L)\,,\quad\G(L)\,,\quad\rho(L)$$ its domain, kernel,
range, graph and resolvent set respectively. $\H$ denotes a Hilbert
space with scalar product $\langle\cdot,\cdot\rangle$ and
corresponding norm $\|\cdot\|$; we also make use of an auxiliary Hilbert space $\fh$ with scalar product 
$(\cdot,\cdot)$ and corresponding norm $|\cdot|$.
\par
Given a closed, densely defined, 
symmetric operator $$S:\D(S)\subseteq \H\to \H$$ with equal deficiency indices, by von Neumann's theory one has (here the direct sums are given w.r.t. the graph inner product of $S^{*}$)
$$
\D(S^{*})=\D(S)\oplus \K_{+}\oplus\K_{-}\,, \quad \K_{\pm}:=\K(-S^{*}\pm i)\,,$$
$$S^{*}(\phi_{\circ}\oplus\phi_{+}
\oplus\phi_{-})=S\phi_{\circ}+i\,\phi_{+}-i\,\phi_{-}\,,
$$  
and any self-adjoint extension of $S$ is of the kind $A_{U}=S^{*}|\G(U)$,  the 
restriction of $S^{*}$ to $\G(U)$, where $U:\K_{+}\to\K_{-}$ is unitary.  Therefore, fixing a unitary $U_{\circ}$ and 
posing $A :=A_{U_{\circ}}$, one has 
$$S=A |\K(\tau_{\circ})\,,\quad \tau_{\circ}:\D(A )\to\fh_{\circ}\,,
$$ 
where 
$$\fh_{\circ}=\K_{+}\,,\quad\tau_{\circ}=P_{+}\,,$$ and $P_{+}$ is the orthogonal (w.r.t. the graph inner product of $S^{*}$) projection onto $\K_{+}$. Since $\K(\tau_{\circ})=\K(\tau)$ where $\tau=M\tau_{\circ}$ and $M:\fh_{\circ}\to\fh$ is any continuous linear bijection, in the search of  the self-adjoint extension of $S$, we can  consider the following equivalent problem: determine all the self-adjoint extensions of $A|\K(\tau)$, 
where
$$
\tau:\D(A)\to\fh 
$$
is a linear, continuous (with respect to the graph norm on $\D(A )$), surjective  map onto 
an auxiliary Hilbert space $\fh$ with its kernel $\K(\tau)$ dense in $\H$. Typically $A $ is a differential operator, $\tau$ 
is some trace (restriction) operator along a null subset $N$ and $\fh$ is some function space over $N$. \par
We suppose that the spectrum of $A $ does not coincide with the whole real line and so, by eventually adding a constant to $A $, we make the following hypothesis:
$$0\in\rho(A )\,.$$
By the results provided in \cite{[P08]} and \cite{[P01]} (to which we refer for proofs and  connections with equivalent formulations, in particular with boundary triplets theory) one has the following
\begin{theorem}\label{estensioni} 
\par\noindent The set of all self-adjoint extensions of $S$ is parametrized by the set $\E(\fh)$ of couples $(\Pi,\Theta)$, where $\Pi$ is an orthogonal projection in $\fh$ and  $\Theta$ is a
self-adjoint operator in $\R(\Pi)$. If $A^{\Pi,\Theta}  $ denotes the self-adjoint extension corresponding to $(\Pi,\Theta)\in \E(\fh)$ 
 then 
\begin{equation*}
A^{\Pi,\Theta}  :\D(A^{\Pi,\Theta}  )\subseteq\H\to\H\,,\quad
A^{\Pi,\Theta}  \phi:=A \phi_0\,,
\end{equation*}
\begin{align*}
\D(A^{\Pi,\Theta}  )
:=\left\{\phi=\phi_0+G_0
\xi_\phi\,,\ \phi_0\in \D(A )\,,\, \xi_\phi\in
\D(\Theta)\,,\,\Pi\tau
\phi_0=\Theta\xi_\phi
\right\}\,,
\end{align*}
where 
$$
G_z:\fh\to\H\,,\quad G_z:=\left(\tau(-A +\bar z)^{-1}\right)^*\,,\quad z\in\rho(A )\,.
$$
Moreover the resolvent of $A^{\Pi,\Theta}  $ is given, for any 
$z\in\rho(A )\cap\rho(A^{\Pi,\Theta}  )$, by the Kre\u\i n's type formula
$$
(-A^{\Pi,\Theta}  +z)^{-1}=(-A +z)^{-1}+G_z
\Pi(\Theta+z\Pi G_0^*G_z\Pi)^{-1}\Pi G^*_{\bar z}\,.
$$
\end{theorem}
\begin{remark}
Notice that the extension corresponding to $\Pi={\mathsf 0}$ is $A $ itself. The extension corresponding to $(\uno,\Theta)$ is denoted by $A^{\Theta}$ and 
everywhere we omit the index $\Pi$ in the case $\Pi=\uno$. 
By \cite{[P04]}, Corollary 3.2, the sub-family $\{A^{\Theta}: \Theta\ \text{self-adjoint}\}$ gives all singular perturbations of $A$, where we say that $\hat A$ is a singular perturbation of $A$  whenever the set $\{\phi\in \D(A)\cap\D(\hat A): A\phi=\hat A\phi\}$ is dense in $\H$ (see \cite{[Ko]}).
\end{remark}
\begin{remark}
The operator $G_z$ is injective (by surjectivity of $\tau$) and for any $z\in\rho(A )$ one has (see \cite{[P01]}, Remark 2.8)
\begin{equation}\label{1.1}
\R(G_z)\cap\D(A )=\{0\}\,,
\end{equation}
so 
that the decomposition 
appearing in  $\D(A^{\Pi,\Theta}  )$ is unique. Moreover (see \cite{[P01]}, Lemma 2.1)
\begin{equation}\label{1.2}
G_{w}-G_{z}=(z-w)(-A+w)^{-1}G_{z}\,.
\end{equation}
\end{remark}
\end{section}
\begin{section}{The common point spectrum} 
Given a self-adjoint operator $A$ let us denote by 
$$
\sigma(A)\,,\quad \sigma_{p}(A)\,,\quad \sigma_{d}(A)
$$
its full, point and  discrete spectrum respectively.\par
Given $\lambda\in\sigma_p(A )$, we denote by $P_\lambda$ the orthogonal
projector onto the corresponding eigenspace 
$\H_\lambda\subseteq\D(A )$ and pose $P_\lambda^\perp:=\uno-P_\lambda$. \par 
Given $\lambda\in\sigma_p(A^{\Pi,\Theta}  )$, we denote by 
$\H_\lambda^{\Pi,\Theta} \subseteq\D(A^{\Pi,\Theta}  )$ the corresponding eigenspace.
\par
As regards the eigenvalues of $A^{\Pi,\Theta}  $ which are not in the spectrum of $A $  a complete answer is given by the following result which is consequence of Kre\u\i n's resolvent formula (see \cite{[DM]}, Section 2, Propositions 1 and 2,  and \cite{[P04]}, Theorem 3.4):
\begin{lemma} $$\lambda\in
\rho(A ) \cap \sigma_p(A^{\Pi,\Theta}  )\quad\iff\quad 
0\in\sigma_p(\Theta+\lambda\Pi G_0^*G_\lambda\Pi)\,,$$
$$
\H_{\lambda}^{\Pi,\Theta}=\{G_\lambda\xi\,,\ \xi\in\K(\Theta+\lambda\Pi G_0^*G_\lambda\Pi)\}\,.
$$
\end{lemma}
Here we are interested in the common eigenvalues, i.e. in the points in 
$\sigma_{p}(A )\cap \sigma_{p}(A^{\Pi,\Theta}  )$. Therefore we take $\lambda\in\sigma_{p}(A )$ and we  
look for solutions $\phi\in \D(A^{\Pi,\Theta}  )$ of the eigenvalue equation
$$
A^{\Pi,\Theta} \phi=\lambda\phi\,,
$$
i.e., by Theorem \ref{estensioni},  
$$
(A -\lambda)\phi_0=\lambda G_0\xi_\phi\,.
$$
By
$$
(A -\lambda)P_{\lambda}\phi_{0}=\zero\,,\qquad 
(A -\lambda)P^{\perp}_{\lambda}\phi_{0}\in\R(P^{\perp}_{\lambda})\,,
$$
this is equivalent to the couple of equations
\begin{equation}\label{eq1}
P_\lambda G_0\xi_\phi=0\,,
\end{equation}
\begin{equation}\label{eq2}
(A -\lambda)P^\perp_\lambda\phi_0=\lambda
P_\lambda^\perp G_0\xi_\phi
\end{equation}
together with the constraint  
\begin{equation}\label{eq3}
\xi_\phi\in\D(\Theta)\subseteq\R(\Pi)\,,\quad\Pi\tau\phi_0=\Theta\xi_\phi\,.
\end{equation}
Equation (\ref{eq1}) gives, for all $\psi\in \H$,
$$
0=\langle G_0\xi_\phi, P_\lambda\psi\rangle=-\langle\xi_\phi,\tau A^{-1}P_\lambda\psi\rangle=-\frac{1}{\lambda}\,\langle\xi_\phi,\tau P_\lambda\psi\rangle
$$
and so
$$
\xi_\phi\in (\R(\tau P_\lambda))^\perp\,.
$$
If $\R(\Pi)\cap(\R(\tau P_\lambda))^\perp=\{0\}$ then, since $G_0$ is injective, one has that in this case $\phi$ is an eigenvector with eigenvalue $\lambda$ 
if and only if $\phi\in\H_\lambda$ and $\Pi\tau\phi=0$.\par
Conversely suppose  that 
$\R(\Pi)\cap(\R(\tau P_\lambda))^\perp\not=\{0\}$ and moreover that $\lambda$ is an isolated eigenvalue. Then $\lambda\in \rho(A |\H_\lambda^\perp)$ and (\ref{eq2}) gives
$$
P_\lambda^\perp\phi_0=-\lambda(-A +\lambda)^{-1} P_\lambda^\perp G_0\xi_\phi\,.
$$
By $\Pi\tau\phi_0=\Theta\xi_\phi$ then one gets
\begin{align}\label{eq4.1}
\Pi\tau P_\lambda\phi_0
=&\Theta\xi_\phi-\Pi\tau P_\lambda^\perp\phi_0
=(\Theta+\lambda\Pi\tau(-A +\lambda)^{-1} 
P_\lambda^\perp G_0\Pi)\xi\,.
\end{align}
By defining 
$$
G_\lambda^\perp:\fh\to \H\,,\quad G_\lambda^\perp
:=(\tau(-A +\lambda)^{-1}P_\lambda^\perp)^*\,,
$$
and by $(G_\lambda^\perp)^* G_0=G_0^*G_\lambda^\perp$ (this relation is consequence of \eqref{1.2}), (\ref{eq4.1}) is equivalent to 
\begin{equation*}
\Pi\tau P_\lambda\phi_0=(\Theta+\lambda\Pi G_0^*G_\lambda^\perp\Pi)\xi
\,.
\end{equation*}
Moreover by $(-A +\lambda)^{-1}P_\lambda^\perp
G_0=-A ^{-1}G_\lambda^\perp$ one has
\begin{align*}
P_\lambda\phi_0+P_\lambda^\perp\phi_0+G_0\xi_\phi=&
P_\lambda\phi_0+(-\lambda(-A +\lambda)^{-1}P_\lambda^\perp+P_\lambda^\perp)G_0\xi_\phi\\
=&P_\lambda\phi_0
-A (-A +\lambda)^{-1} P_\lambda^\perp G_0\xi_\phi\\
=&
P_\lambda\phi_0+G_\lambda^\perp\xi_\phi\,.
\end{align*}
In conclusion we have proven the following 
\begin{theorem}\label{point spectrum} 
Let $\lambda\in\sigma_p(A )$.
\par\noindent
1) Suppose $$
\R(\Pi)\cap(\R(\tau P_\lambda))^\perp=\{0\}\,.
$$ 
and pose
$$\K^\Pi_\lambda:=\{\psi\in\H_\lambda\,:\,\Pi\tau\psi=0\}\,.
$$
Then
$$
\lambda\in\sigma_p(A^{\Pi,\Theta})\quad\iff\quad 
\K_\lambda^\Pi\not=\{0\}\,;
$$
and 
$$\H_\lambda^{\Pi,\Theta} =
\K_\lambda^\Pi\,.
$$
\vskip10pt\noindent
2) Suppose 
$$
\R(\Pi)\cap(\R(\tau P_\lambda))^\perp\not=\{0\}\,.
$$
and let $\lambda$ be isolated. \par 
Let $\N^{\Pi,\Theta}_\lambda$ be the set 
of couples $(\psi,\xi)\in \H_\lambda\oplus\R(\Pi)$ 
such that
\begin{equation}\label{uno}\xi\in D(\Theta)\cap(\R(\tau P_\lambda))^\perp\,,
\end{equation} 
\begin{equation}\label{due}
\Pi\tau \psi=(\Theta+\lambda\Pi G_0^*G_\lambda^\perp\Pi)\,\xi
\,.
\end{equation}
Then
$$\lambda\in\sigma_p(A^{\Pi,\Theta}  )\quad\iff\quad 
\N^{\Pi,\Theta}_\lambda\not=\{0\}\,;
$$
$$\text{\rm dim}(\H^{\Pi,\Theta}_\lambda)=\text{\rm dim}(\N^{\Pi,\Theta}_\lambda)$$
and
$$\H^{\Pi,\Theta}_\lambda=
\{\phi\in\H\,:\,\phi=\psi+G_\lambda^\perp\xi\,,\quad (\psi,\xi)\in \N^{\Pi,\Theta}_\lambda\}\,.
$$
\end{theorem}
\begin{remark}\label{rem2} Notice that
$$
(\R(\Pi))^{\perp}\cap\R(\tau P_{\lambda})\not=\{0\}\quad\Longrightarrow\quad\K^{\Pi}_{\lambda}\not=\{0\}\,.
$$
\end{remark}
\begin{remark} Suppose $\lambda\in\sigma_{p}(A)$ is isolated. Noticing that 
$$
\K^{\Pi}_{\lambda}\oplus(\R(\Pi)\cap(\R(\tau P_{\lambda}))^{\perp}\cap\K(\Theta+\lambda\Pi G^{*}_{0}G^{\perp}_{\lambda}\Pi))
\subseteq \N^{\Pi,\Theta}_{\lambda}\,,
$$
one has
$$
\K^{\Pi}_{\lambda}\not=\{0\}\quad\Longrightarrow\quad \lambda\in\sigma_{p}(A^{\Pi,\Theta})\,.
$$
In particular, in the case  $\lambda$ is simple with eigenvector 
$\psi_{\lambda}$, 
$$
\Pi\tau\psi_{\lambda}=0\quad\Longrightarrow\quad \lambda\in\sigma_{p}(A^{\Pi,\Theta})\,.
$$ 
\end{remark}
\begin{remark}\label{rem} Suppose $\R(\tau P_{\lambda})=\{0\}$. Then $\K^{\Pi}_{\lambda}=\H_{\lambda}$ and so, in case $\lambda\in\sigma_{p}(A)$ is isolated,  $\lambda\in\sigma_{p}(A^{\Pi,\Theta})$ and 
$$
\H^{\Pi,\Theta}_{\lambda}=\{\phi=\psi_{\lambda}+G^{\perp}_{\lambda}\xi\,,\ 
\psi_{\lambda}\in\H_{\lambda}\,,\ \xi\in\R(\Pi)\cap\K(\Theta+\lambda \Pi G_{0}^{*}G_{\lambda}^{\perp}\Pi)\}\,.
$$
\end{remark}
\begin{remark} The papers \cite{[BGP]} and \cite{[MS]} contain results related to the ones given by Theorem \ref{point spectrum} (see Theorem 3.6 in  \cite{[MS]} and Theorem 4.7 in \cite{[BGP]}). We thank Konstantin Pankrashkin for the communication. 
\end{remark}
\end{section}
\begin{section}{Examples.}
\begin{subsection}{Rank-one singular perturbations.}\label{R1}
\par
Suppose $\fh=\CO$. Then $\Pi=\uno$, $\Theta=\theta\in\RE$ 
and either $\R(\tau P_\lambda)=\CO$ or $\R(\tau P_\lambda)=\{0\}$.\par 
If $\R(\tau P_\lambda)=\CO$ then $\lambda\in\sigma_p(A^{\theta})$ if and only if $\K_\lambda\not=\{0\}$, where
$$
\K_\lambda:=\{\psi\in \H_\lambda\,:\,
\tau\psi=0\}\,.
$$ 
Since $\R(\tau P_\lambda)=\{0\}$ if and only if $\K_\lambda=\H_\lambda$, when $\lambda$ is isolated and $\R(\tau P_\lambda)=\{0\}$ one has
$$
\N^{\theta}_\lambda=\K_\lambda\oplus
\{\xi\in\CO\,:\,(\theta+\lambda\langle G_0,G_\lambda^\perp\rangle)\,\xi=0\}
$$
and so
$$
\theta+\lambda\langle
G_0,G_\lambda^\perp\rangle=0\quad\Longrightarrow\quad 
\N^{\theta}_\lambda=
\K_\lambda\oplus\CO\equiv\H_\lambda\oplus\CO\,,
$$
$$
\theta+\lambda\langle
G_0,G_\lambda^\perp\rangle\not=0\quad\Longrightarrow\quad 
\N^{\theta}_\lambda=\K_\lambda\oplus\{0\}\equiv\H_\lambda \,.
$$
In conclusion when $\fh=\CO$ and $\lambda\in \sigma_p(A)$ is isolated, 
\begin{equation}\label{spR1}
\lambda\in\sigma_p(A^{\theta})\iff
\K_\lambda\not=\{0\}
\end{equation} 
and
$$
\H^{\theta}_{\lambda}=
\{\psi=\psi_{\lambda}+G_{\lambda}^{\perp}\xi\,,\ \psi_{\lambda}\in\H_{\lambda}\,,\ (\theta+\lambda\langle
G_0,G_\lambda^\perp\rangle)\xi=0\}\,.
$$
In particular if $\lambda$ is
a simple isolated eigenvalue of $A $ with corresponding eigenfunction 
$\psi_\lambda$, then $\lambda\in\sigma_p(A^{\theta})$ if and only if 
$\tau\psi_\lambda=0$. For example, if $\H=L^2(\Omega)$  and $\tau:\D(A )\to\CO$ is the evaluation map at $y\in\Omega$, $\tau\psi:=\psi(y)$, then $\lambda$ is preserved if and only if $y$ belongs to the nodal set (if any) of $\psi_\lambda$. Thus if $A $ is (minus) the  Dirichlet  Laplacian on a bounded open set $\Omega\subset\RE^{d}$, $d\le 3$, its lowest eigenvalue is never preserved under a point perturbation. Analogous results hold in the case $A $ is the Laplace-Beltrami operator on a compact $d$-dimensional Riemannian manifold $M$, $d\le 3$, thus reproducing the ones given in \cite{[DV]}, Theoreme 2, part 1.
 \end{subsection}
\begin{subsection}{\bf The {\v S}eba billiard.} \par Let 
$$A=\Delta:\D(A)\subset L^2(R)\to L^2(R)\,,$$ 
$$
\D(A)=\{\phi\in C(\overline R): \Delta\phi\in L^{2}(R)\,,\ \phi(\mathsf x)=0\,,\ {\mathsf x}\in\partial R\}\,,
$$
be the Dirichlet Laplacian on the rectangle 
$R=(0,a)\times (0,b)$. Then $$\sigma(A)=\sigma_d(A)=\left\{\lambda_{m,n}\,,\ (m, n)\in{\mathbb N}^{2}\right\}$$ and 
$$\H_{\lambda_{m,n}}=\text{\rm span}\{\psi_{m',n'}\,:\,\lambda_{m',n'}=\lambda_{m,n}\}\,,$$
where
$$\lambda_{m,n}:=-\pi^2\left(\frac{m^2}{a^2}+\frac{n^2}{b^2}\right)
$$
and
$$
\psi_{m,n}(\mathsf x):=\sin\left(\frac{m\pi x_{1}}{a}\right)\,
\sin\left(\frac{n\pi x_{2}}{b}\right)\,,\quad {\mathsf x}\equiv(x_{1},x_{2})\,.
$$
Let $$\tau\psi:=\psi(\mathsf y)\,,$$
so that $A^{\theta}$ describes a ``{\v S}eba billiard'', i.e. the Dirichlet Laplacian on the rectangle $R$ with a point perturbation placed at the point $\mathsf y\equiv(y_{1},y_{2})$ (see \cite{[S]}).\par  
Since $\sigma(A)=\sigma_d(A)$, by the invariance of the essential spectrum under finite rank perturbations, one 
has $\sigma(A^\theta)=\sigma_d(A^\theta)$ and, by \eqref{spR1}, $\lambda_{m,n}\in\sigma(A)\cap\sigma(A^{\theta})$ if and only $$
\forall\, (m',n')\ \text{s.t.}\ \lambda_{m',n'}=\lambda_{m,n}\,,\quad \sin\left(\frac{m'\pi y_{1}}{a}\right)\,\sin\left(\frac{n'\pi y_2}{b}\right)=0\,.
$$
Equivalently  
$$\sigma(A)\cap\sigma(A^{\theta})=\emptyset\quad\iff\quad
\left(\frac{y_{1}}{a},\frac{y_2}{b}\right)\notin{\mathbb Q}^{2}\,.
$$ 
If there exists relatively prime integers $1\le p<q$ such that $\frac{y_{1}}{a}=\frac p q$ while $\frac{y_2}{b}$ is irrational, then $$ \sigma(A)\cap\sigma(A^\theta)=\{\lambda_{kq,n}\,,\ 
(k, n)\in{\mathbb N}^{2}\}\,.
$$
Analogously if 
$\frac{y_{1}}{a}$ is irrational and $\frac{y_2}{b}=\frac p q$ then $$ \sigma(A)\cap\sigma(A_\theta)=\{\lambda_{m,kq}\,,\ (m,k)\in{\mathbb N}^{2}\}$$
while if $\frac{y_{1}}{a}=\frac p q$ and $\frac{y_2}{b}=\frac r s$, then $$ \sigma(A)\cap\sigma(A^\theta)=
\{\lambda_{kq,n}\,,\ (k, n)\in{\mathbb N}^{2}\}\cup\{\lambda_{m,ks}\,,\ (m, k)\in{\mathbb N}^{2}\}\,.$$
\end{subsection}
\begin{subsection}{Rank-two singular perturbations}\label{R2}
Let $\fh=\CO^{2}$. Then either $\Pi=\uno$ or $\Pi=w\otimes w$, $w\in\CO^{2}$, $|w|=1$. Let $\lambda\in\sigma_{p}(A)$. \par
1.1) $\R(\tau P_{\lambda})=\CO^{2}$, $\Pi=\uno$. Then $\lambda\in\sigma_{p}(A^{\Theta})$ if and only if there exists 
$\psi\in\H_{\lambda}\backslash\{0\}$ such that $\tau\psi=0$.\par
1.2) $\R(\tau P_{\lambda})=\CO^{2}$, $\Pi=w\otimes w$. Then $\lambda\in\sigma_{p}(A^{\Pi,\Theta})$ if and only if there exists 
$\psi\in\H_{\lambda}\backslash\{0\}$ such that $w\cdot\tau\psi=0$.
\par
Now suppose further that $\lambda\in\sigma_{p}(A)$ is isolated.\par
2.1) $\R(\tau P_{\lambda})=\text{span}(\xi_{\lambda})\simeq\CO$, $|\xi_{\lambda}|=1$, $\Pi=\uno$. Decomposing equation \eqref{due} w.r.t. the orthonormal base $\{\xi_{\lambda},\xi_{\lambda}^{\perp}\}$ one gets that $\N_{\lambda}^{\Theta}\not=\{0\}$ if and only if there exists $\zeta\equiv(\zeta_{1},\zeta_{2})\in\CO^{2}\backslash\{0\}$ solving
$$
\begin{cases}
\zeta_{1}=(\xi_{\lambda}\cdot (\Theta+\lambda G_0^{*}G_{\lambda}^{\perp})\xi^{\perp}_{\lambda})\zeta_{2}\\
0=(\xi_{\lambda}^{\perp}\cdot (\Theta+\lambda G_0^{*}G_{\lambda}^{\perp})\xi^{\perp}_{\lambda})\zeta_{2}\,.
\end{cases}
$$
Hence 
$$
\lambda\in\sigma_{p}(A^{\Theta})\iff (\xi_{\lambda}^{\perp}\cdot (\Theta+\lambda G_0^{*}G_{\lambda}^{\perp})\xi^{\perp}_{\lambda})=0\,.
$$
2.2) $\R(\tau P_{\lambda})=\text{span}(\xi_{\lambda})\simeq\CO$, $\Pi=w\otimes w$. Let us use the decomposition $w=w_{||}+w_{\perp}$ w.r.t. the orthonormal base $\{\xi_{\lambda},\xi_{\lambda}^{\perp}\}$.  If $w_{||}=0$ then 
$\K^{\Pi}_{\lambda}\not=\{0\}$ and so $\lambda\in\sigma_{p}(A^{\Pi,\Theta})$. If  $w_{||}\not=0$ then $\K^{\Pi}_{\lambda}=\{0\}$ and $\R(\Pi)\cap (\R(\tau P_{\lambda}))^{\perp}=\{0\}$, 
thus $\lambda\notin\sigma_{p}(A^{\Pi,\Theta})$. In conclusion
$$
\lambda\in\sigma_{p}(A^{\Pi,\Theta})\quad\iff\quad w=\xi_{\lambda}^{\perp}\,.
$$
3) $\R(\tau P_{\lambda})=\{0\}$. In this case  $\lambda\in\sigma_{p}(A^{\Pi,\Theta})$.
\end{subsection}
\begin{subsection} {The Laplacian on a bounded interval} \label{Lap} Let 
$$A:\D(A)\subseteq L^2(0,a)\to
L^2(0,a)\,,\qquad A\phi=\phi''\,,$$ 
$$\D(A)=\{\phi\in C^{1}[0,a]\,:\,\phi''\in L^{2}(0,a)\,,\ \phi(0)=\phi(a)=0\}$$
be the Dirichlet Laplacian on the bounded interval $(0,a)$
and pose
$$
\tau:\D(A)\to\CO^2\,,\qquad \tau\phi\equiv\gamma_{1}\phi:=\left(\phi'(0),-\phi'(a)\right)\,.
$$
Therefore $S=A|\K(\tau)$ is the minimal Laplacian with domain 
\begin{align*}
&\D(S)\\
=&\{\phi\in C^{1}[0,a]\,:\,\phi''\in L^{2}(0,a)\,, \ 
\phi(0)=\phi'(0)=\phi(a)=\phi'(a)=0\}\,.
\end{align*}
and the self-adjoint extensions of $S$ are rank-two perturbations of the Dirichlet Laplacian $A$.
One has $$\sigma(A)=\sigma_{d}(A)=\{\lambda_{n}\}_{1}^{\infty}\,,\quad \lambda_{n}=-\left(\frac{n\pi}{a}\right)^2$$ and the normalized 
eigenvector corresponding to $\lambda_{n}$ is  $$\psi_{n}(x)=\sqrt{\frac 2a}\,\sin\left(\frac{n\pi x}{a}\right)\,.$$ By Theorem \ref{estensioni} and by the change of extension parameter (here $P_{0}$ represents the Dirichlet-to-Neumann operator)
$$(\Pi,\Theta)\mapsto(\Pi,B)\,,\quad
B:=\Theta-\Pi P_{0}\Pi\,,\quad P_{0}\equiv
\frac1a\left(\begin{matrix}\,\,\ 1&-1\\-1&\,\,\ 1\end{matrix}\right)
$$
any self-adjoint extension of the minimal Laplacian $S$ is of the kind $A^{\Pi,B}$, $(\Pi,B)\in\E(\CO^{2})$, where 
$$
A^{\Pi,B}:\D(A^{\Pi,B})\subset L^{2}(0,a)\to L^{2}(0,a)\,,\quad A^{\Pi,B}\phi=\phi''\,,
$$
$$
\D(A^{\Pi,B})=\{\phi\in C^{1}[0,a]:\phi''\in L^{2}(0,a)\,,\ \gamma_{0}\phi\in\R(\Pi)\,,\ 
\Pi \gamma_{1} \phi=B\gamma_{0}\phi\}\,,
$$
(see e.g. \cite{[P08]}, Example 5.1). Here $\gamma_{0}\phi:=(\phi(0),\phi(a))$. \par 
The case $\Pi=0$ reproduces $A$ itself, the case $\Pi=\uno$, $B=\left(\begin{matrix}b_{11}&b_{12}\\\bar b_{12}&b_{22}\end{matrix}\right)$, $b_{11},b_{22}\in\RE$, $b_{12}\in\CO$, gives
the boundary conditions 
$$
\begin{cases}
b_{11}\,\phi(0)-\phi'(0)+b_{12}\,\phi(a)=0\,, 
\\ \bar b_{12}\,\phi(0)+b_{22}\,\phi(a)+\phi'(a)=0\,,
\end{cases}
$$
and the case $\Pi=w\otimes w$, $w\equiv(w_{1},w_{2})\in\CO^{2}$, $|w_{1}|^{2}+|w_{2}|^{2}=1$, $B\equiv b\in\RE$, gives the boundary conditions 
$$
\begin{cases}
w_{2}\,\phi(0)-w_{1}\,\phi(a)=0\,,\\
\bar w_{1}\,(b\,\phi(0)-\phi'(0))
+\bar w_{2}\,(b\,\phi(a)+\phi'(a))=0\,.
\end{cases}
$$
By the invariance of the essential spectrum under finite rank perturbations, $\sigma(A^{\Pi,B})=\sigma_{d}(A^{\Pi,B})$. Now we use the results given in subsection \ref{R2}. 
One has 
$$\R(\tau P_{\lambda_{n}})=\text{span}(\hat \xi_n )\,,\quad \hat \xi_n \equiv\frac1{\sqrt 2}\,\left(1,(-1)^{n-1}\right)\,.
$$
Let $\Pi=\uno$ and  $\hat \xi_n ^{\perp}\equiv\frac1{\sqrt 2}\,\left(1,(-1)^{n}\right)$. By point 2.1 in subsection \ref{R2} we known that $\lambda_{n}\in\sigma(A^{B})$ if and only if 
$\hat \xi_n ^{\perp}\cdot(B+P_{0}+\lambda_{n} G_{0}^{*}G_{\lambda_{n}}^{\perp})\hat \xi_n ^{\perp}=0$. Since the resolvent of $A$ is explicitly known, $\hat \xi_n ^{\perp}\cdot(B+P_{0}+\lambda_{n} G_{0}^{*}G_{\lambda_{n}}^{\perp})\hat \xi_n ^{\perp}$ can be calculated. 
However we use here a short cut which avoids any calculation: the Neumann Laplacian corresponds to $B=0$ and we know that its spectrum is  $\{0\}\cup\sigma(A)$, thus 
\begin{equation}\label{perp}
\hat \xi_n ^{\perp}\cdot (P_{0}+ \lambda_{n} G_{0}^{*}G_{\lambda_{n}}^{\perp})\hat \xi_n ^{\perp}=0
\end{equation}
Therefore we obtain
$$
\lambda_{n}\in\sigma(A^{B})\quad\iff\quad b_{11}+b_{22}+2\,(-1)^{n}\text{Re}(b_{12})=0\,.
$$
If $\Pi=w\otimes w$ by point 2.2 in subsection \ref{R2} one has
$$
\lambda_{n}\in\sigma(A^{\Pi,B})\quad\iff\quad w=\hat\xi_{n}^{\perp}\,.
$$
In both cases 
$$
\lambda_{n}\in\sigma(A^{\Pi,B})\quad\iff \lambda_{n+2}\in\sigma(A^{\Pi,B})\,.$$
Moreover
$$
\sigma(A)\subseteq\sigma(A^{\Pi,B})\quad\iff\quad \text{$\Pi=\uno$ and $b_{11}+b_{22}=0$, Re$(b_{12})=0$}\,.
$$ 
\end{subsection}
\begin{subsection}{\bf Equilateral Quantum Graphs.}\par
Let $\H=\oplus_{k=1}^N L^2(0,a)$ and $A_{N}=\oplus_{k=1}^{N} A$, where $A$ is defined as in subsection \ref{Lap} (to which we refer for notations). Then $\sigma(A_{N})=\sigma_d(A_{N})
=\sigma(A)$ and the eigenfunctions corresponding to the $N$-fold degenerate eigenvalue $\lambda_{n}$ are
$$\Psi_{k,n}=\oplus_{i=1}^N\psi_{i,k,n}\,,\ k=1,\dots, N\,,\quad 
\psi_{i,k,n}=\begin{cases}0\,,&i\not=k\\\psi_{n}\,,&i=k\,.\end{cases}$$  
By taking 
$$
\tau:\D(A_{N})\equiv\oplus_{k=1}^{N}\D(A)\to\oplus_{k=1}^{N}\CO^{2}\equiv\CO^{2N}\,,\quad \tau=\oplus_{k=1}^N\gamma_{1}\,,
$$
one gets, by Theorem \ref{estensioni}, self-adjoint extensions describing 
quantum graphs (see e.g. \cite{[ku]}) with $N$ edges of the same length $a$ .  By Theorem \ref{estensioni} and by the change of extension parameter 
$$(\Pi,\Theta)\mapsto(\Pi,B)\,,\quad
B:=\Theta-\Pi (\oplus_{k=1}^{N}P_{0})\Pi\,,
$$
such extensions are of the kind $A^{\Pi,B}$, $(\Pi,B)\in\E(\CO^{2N})$, where (see \cite{[P08]}, Example 5.2). 
$$
A^{\Pi,B}:\D(A^{\Pi,B})\subset \oplus_{k=1}^{N}L^{2}(0,a)\to \oplus_{k=1}^{N}L^{2}(0,a)\,,\quad 
$$
$$
A^{\Pi,B}(\oplus_{k=1}^{N}\phi_{k})=\oplus_{k=1}^{N}\phi_{k}''\,,
$$
\begin{align*}
&\D(A^{\Pi,B})=\{\oplus_{k=1}^{N}\phi_{k}:\phi_{k}\in C^{1}[0,a]\,,\ \phi_{k}''\in L^{2}(0,a)\,,\\ &(\oplus_{k=1}^{N}\gamma_{0}\phi_{k})\in\R(\Pi)\,,\ 
\Pi (\oplus_{k=1}^{N}\gamma_{1} \phi_{k})=B(\oplus_{k=1}^{N}\gamma_{0}\phi_{k})\}\,.
\end{align*}
The couple $(\Pi,B)$ represents the connectivity of the quantum graph.\par
1) $\Pi=\uno$. Given $\lambda_{n}\in\sigma(A)$, we pose 
$$
\CO^{2N}_{||}:=\oplus_{k=1}^{N}\text{span}(\hat \xi_{n})\simeq\CO^{N}\,,\quad 
\CO^{2N}_{\perp}:=\oplus_{k=1}^{N}\text{span}(\hat \xi_{n}^{\perp})\simeq\CO^{N}\,,
$$
so that $\R(\tau P_{\lambda_{n}})=\CO^{2N}_{||}$,  $(\R(\tau P_{\lambda_{n}}))^{\perp}=\CO^{2N}_{||}$, $\CO^{2N}=\CO^{2N}_{||}\oplus \CO^{2N}_{\perp}$ and for any linear operator $L:\CO^{2N}\to\CO^{2N}$ we can consider the block decomposition $L=\left(\begin{matrix}L_{||}&L_{||\perp}\\ (L_{||\perp})^{*}&L_{\perp}\end{matrix}\right)$. By using such decompositions in equation \eqref{due} one gets that 
$\N_{\lambda_{n}}^{\Theta}\not=\{0\}$, $\Theta=B+\oplus_{k=1}^{N}P_{0}$, if and only if there exists $\zeta\not=\{0\}$, $\zeta=\zeta_{||}\oplus\zeta_{\perp}\in \CO_{||}^{2N}\oplus\CO^{2N}_{\perp}$ solving
$$
\begin{cases}
\zeta_{||}=(B+\oplus_{k=1}^{N}P_{0}+\lambda_{n} G_0^{*}G_{\lambda_{n}}^{\perp})_{||\perp}\,\zeta_{\perp}\\
0=(B+\oplus_{k=1}^{N}P_{0}+\lambda_{n} G_0^{*}G_{\lambda_{n}}^{\perp})_{\perp}\,\zeta_{\perp}\,.
\end{cases}
$$
By \eqref{perp} one obtains $(\oplus_{k=1}^{N}P_{0}+\lambda_{n} G_0^{*}G_{\lambda_{n}}^{\perp})_{\perp}=0$. Therefore one gets
$$
\lambda_{n}\in\sigma(A^{B})\quad\iff\quad 
\det(B_{\perp})=0\,.
$$
2) $\Pi\not=\uno$. Given $\lambda_{n}\in\sigma(A)$ we pose 
$$
\R(\Pi)_{||}:=\R(\Pi)\cap(\oplus_{k=1}^{N}\text{span}(\hat \xi_{n}))\quad 
\R(\Pi)_{\perp}:=\R(\Pi)\cap(\oplus_{k=1}^{N}\text{span}(\hat \xi_{n}^{\perp}))\,,
$$
so that $\R(\Pi)\cap \R(\tau P_{\lambda_{n}})=\R(\Pi)_{||}$,  
$\R(\Pi)\cap(\R(\tau P_{\lambda_{n}}))^{\perp}=\R(\Pi)_{\perp}$, 
$\R(\Pi)=\R(\Pi)_{||}\oplus \R(\Pi)_{\perp}$ and for any linear operator 
$L:\R(\Pi)\to \R(\Pi)$ we can consider the block decomposition $L=\left(\begin{matrix}L_{||}&L_{||\perp}\\ (L_{||\perp})^{*}&L_{\perp}\end{matrix}\right)$.\par
Define $\hat\xi_{k,n}=\oplus_{i=1}^{n}\hat\xi_{i,k,n}\in\CO^{2N}$ and $\hat\xi_{k,n}^{\perp}=\oplus_{i=1}^{n}\hat\xi_{i,k,n}^{\perp}\in\CO^{2N}$, $k=1,\dots,N$, by
$$
\hat\xi_{i,k,n}:=\begin{cases}0\,,& i\not=k\\
\hat\xi_{n}\,,& i=k\,,\end{cases}\quad \hat\xi_{i,k,n}^{\perp}:=\begin{cases}0\,,& i\not=k\\
\hat\xi_{n}^{\perp}\,,& i=k\,,\end{cases}$$ 
If $\Pi\hat\xi_{k,n}^{\perp}=0$ for all $k$ then $\R(\Pi)_{\perp}=\{0\}$ and in this case
$$
\lambda\in\sigma_{p}(A^{\Pi,B})\quad\iff\quad \exists k\ \text{s.t}\ \Pi\hat\xi_{k,n}=0\,.
$$
If there exists $k'$ such that $\Pi\hat\xi_{k',n}^{\perp}\not=0$ then 
$\R(\Pi)_{\perp}\not=\{0\}$.  By Remark \ref{rem2}
$$
\exists k\ \text{s.t}\ \Pi\hat\xi_{k,n}=0\quad\Longrightarrow\quad \lambda\in\sigma_{p}(A^{\Pi,B})\,.
$$ 
Suppose now $\Pi\hat\xi_{k,n}\not=0$ for all $k$, i.e. $\K_{\lambda_{n}}^{\Pi}=\{0\}$.
Then, using the above decompositions in equation \eqref{due} one gets that 
$\N_{\lambda_{n}}^{\Pi,\Theta}\not=\{0\}$, $\Theta=B+\Pi(\oplus_{k=1}^{N}P_{0})\Pi$, if and only if there exists $\zeta\not=0$, $\zeta=\zeta_{||}\oplus\zeta_{\perp}\in \R(\Pi)_{||}
\oplus\R(\Pi)_{\perp}$ solving
$$
\begin{cases}
\zeta_{||}=(B+\Pi(\oplus_{k=1}^{N}P_{0}+\lambda_{n} G_0^{*}G_{\lambda_{n}}^{\perp})\Pi)_{||\perp}\,\zeta_{\perp}\\
0=(B+\Pi(\oplus_{k=1}^{N}P_{0}+\lambda_{n} G_0^{*}G_{\lambda_{n}}^{\perp})\Pi)_{\perp}\,\zeta_{\perp}\,.
\end{cases}
$$
By \eqref{perp} one obtains $(\Pi(\oplus_{k=1}^{N}P_{0}+\lambda_{n} G_0^{*}G_{\lambda_{n}}^{\perp})\Pi)_{\perp}=0$. Therefore one gets, in case there exists $k'$ such that $\Pi\hat\xi_{k',n}^{\perp}\not=0$ and $\Pi\hat\xi_{k,n}\not=0$ for all $k$,
$$
\lambda_{n}\in\sigma(A^{\Pi,B})\quad\iff\quad 
\det(B_{\perp})=0\,.
$$
\end{subsection}
\end{section}

\end{document}